\documentclass{moriond}

\usepackage{xspace}

\newcommand{\mH}{\ensuremath{\textrm{m}_\mathrm{H}\xspace}}
\newcommand{\widthH}{\ensuremath{\Gamma_\mathrm{H}\xspace}}
\newcommand{\cHW}{\ensuremath{\mathrm{c}_{H\tilde{W}}\xspace}}
\newcommand{\cHB}{\ensuremath{\mathrm{c}_{H\!\tilde{B}}\xspace}}
\newcommand{\cHWB}{\ensuremath{\mathrm{c}_{H\tilde{W}\!B}\xspace}}
\newcommand{\alphacp}{\ensuremath{\alpha^{\mathrm{H}\tau\tau}\xspace}}
\newcommand{\phicp}{\ensuremath{\phi_{CP}\xspace}}
\newcommand{\fbinv}{fb$^{-1}$ }

\newcommand{\Photo}{\includegraphics[height=35mm]{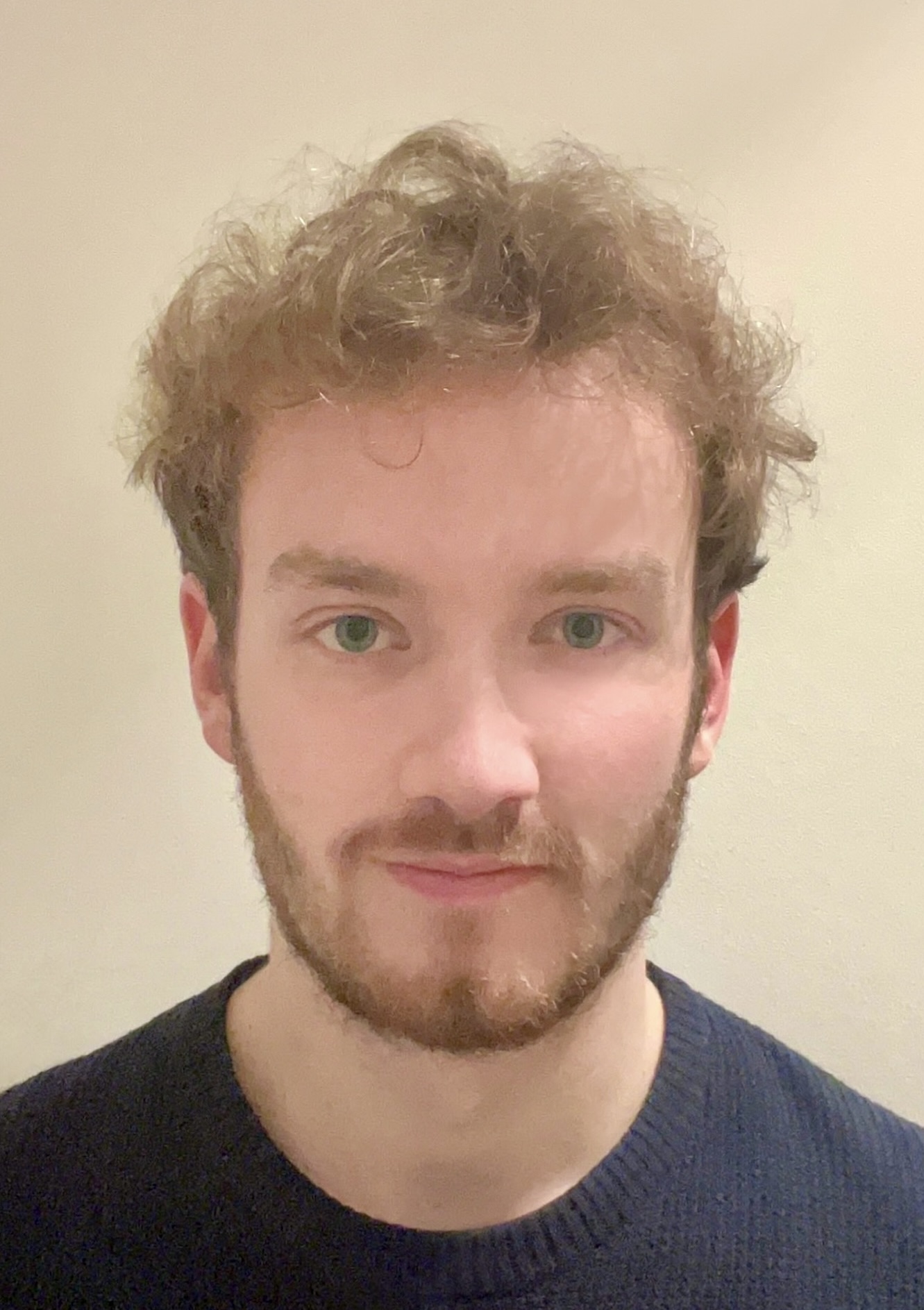}}

\begin{document}
\vspace*{4cm}
\title{Higgs CP studies and other Higgs properties at ATLAS and CMS}

\author{ Lucas Russell \\ on behalf of the ATLAS and CMS Collaborations \footnote{Copyright 2026 CERN for the benefit of the ATLAS and CMS Collaborations. Reproduction of this article or parts of it is allowed as specified in the CC-BY-4.0 license.}}

\address{Imperial College London, London, UK}

\maketitle\abstracts{
Recent measurements of Higgs boson properties using proton-proton collision data recorded at $\sqrt{s}=13$ TeV and $\sqrt{s}=13.6$ TeV with the ATLAS and CMS detectors at the LHC are presented. 
CMS results on the Higgs boson mass and width in the $H\to\gamma\gamma$ channel are reported, together with a measurement of the Higgs boson width in the $H\to WW^*$ final state.
A combined measurement of the charge-parity (CP) nature of the Higgs-vector boson couplings by the ATLAS collaboration, as well as improved results in individual channels by both experiments are summarised.
Finally, the latest CMS analysis of the CP structure of the Higgs-tau lepton Yukawa coupling is reported.
}

\section{Introduction}

The Higgs boson mass, \mH, is not predicted by the standard model (SM) and must be measured experimentally. 
Its precise determination is essential for predicting Higgs couplings, probing electroweak vacuum stability, and testing the consistency of the electroweak sector.
In contrast, the Higgs width, \widthH, is predicted in the SM to be approximately 4.1 MeV. Deviations from this value would indicate beyond the standard model (BSM) physics. However, direct measurements of \widthH\ are challenging due to the larger detector resolution which is of order 1 GeV.
The SM is unable to explain the observed matter-antimatter asymmetry of the universe, with additional sources of charge-parity (CP) violation required. While the Higgs boson is predicted to be CP-even in its interactions with other SM particles, CP-violating couplings could exist in BSM scenarios and can be probed at the LHC.

The latest measurements of Higgs boson properties are reported in these proceedings using proton-proton collision data recorded by the ATLAS and CMS experiments at the CERN LHC during Run 2 (138-140 \fbinv at $\sqrt{s}=13$ TeV) and Run 3 (62-164 \fbinv at $\sqrt{s}=13.6$ TeV).


\section{Higgs boson mass and width measurements}

\subsection{Latest Higgs boson mass measurement}

The Higgs boson mass can be precisely measured in the diphoton decay channel despite the small $H\to\gamma\gamma$ branching ratio of 0.23\%, as it is a relatively clean final state.
The latest measurement of \mH\ has been performed by the CMS collaboration using the Run 2 dataset~\cite{CMS_Hgg_Mass}.
Excellent mass resolution is achieved using precise photon energy calibrations. The energy scale is first measured using electrons from $Z\to ee$ decays, before further corrections are derived using photons from $Z\to\mu\mu\gamma$ decays. Events are categorised using a boosted decision tree, before a fit to the diphoton invariant mass is performed.

\begin{figure}[!htb]
    \centering
    \includegraphics[width=0.67\textwidth]{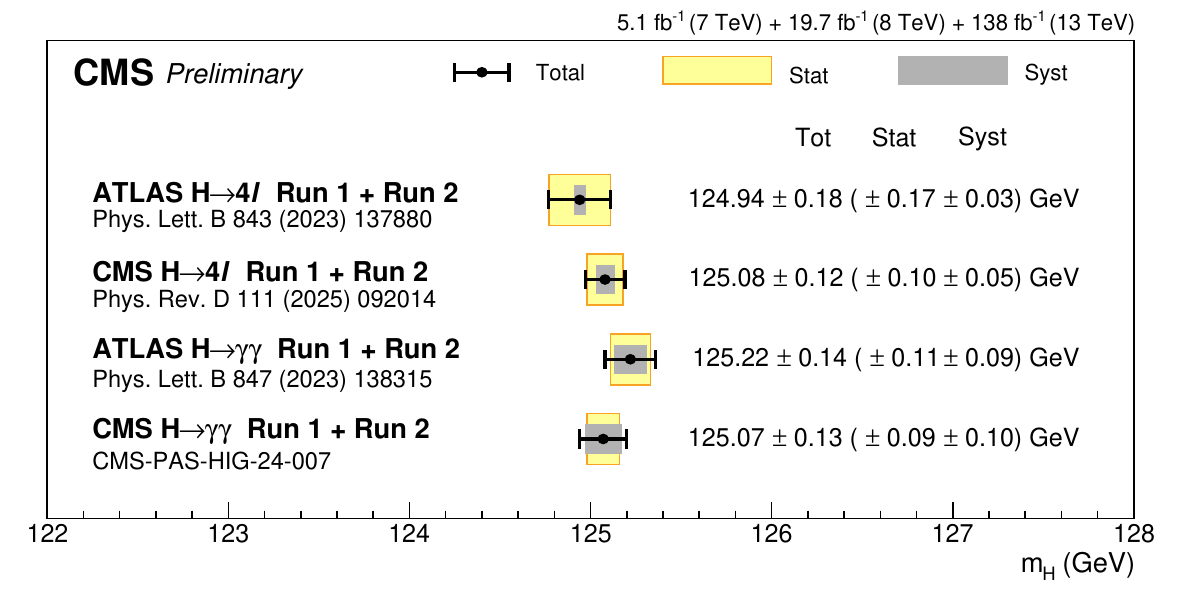}
    \caption{Summary of the ATLAS and CMS Higgs boson mass measurements performed in the  $H\to\gamma\gamma$ and $H\to ZZ^*\to4\ell$ decay channels. The measured values of \mH\ are in good agreement.}
\label{fig:mh}
\end{figure}

The Higgs boson mass is found to be $\mH = 125.14\pm0.11 \textrm{ (syst.)} \pm 0.10 \textrm{ (stat.)}$ GeV. 
The result is combined with the previous CMS measurement using data recorded during LHC Run~1, yielding a combined value of {$\mH = 125.07\pm0.13$ GeV}. 
This result is in good agreement with the CMS result in the $H\to ZZ^*\to4\ell$ decay channel, as well as equivalent measurements by the ATLAS collaboration, as shown in Fig.~\ref{fig:mh}. 

\subsection{Latest Higgs boson width measurements}

The Higgs boson width has been measured by the CMS collaboration from the ratio of off-shell to on-shell $H\to W W^*\to e\nu\mu\nu$ decays using the Run 2 dataset \cite{CMS_HWW_Width}. The on-shell and off-shell events are separated by Higgs boson production mechanism using a deep neural network. The combined off-shell Higgs boson signal strength was found to be $\mu_{\mathrm{off}\textrm{-}\mathrm{shell}} = 1.2^{+0.8}_{-0.7}$. This was used to derive the total decay width $\widthH=3.9^{+2.7}_{-2.2}$ MeV, in agreement with the SM prediction, and complementing measurements in the $H\to ZZ^*$ channel.

A complementary measurement of \widthH\ was performed by CMS with on-shell $H\to\gamma\gamma$ decays using the Run 2 dataset~\cite{CMS_Hgg_Width}. 
The analysis exploits the interference between {$gg\to H \to \gamma\gamma$} signal and the {non-resonant} {$gg\to\gamma\gamma$} background which leads to a \widthH-dependent distortion of the diphoton mass spectrum. The observed limit is $\widthH<92$ MeV at the 95\% confidence level (CL), which represents a large improvement with respect to previous on-shell measurements, while remaining weaker than constraints from on and off-shell comparisons.

\section{Higgs boson CP Properties}

\subsection{Combined measurement of Higgs-vector boson couplings}

A combined measurement of the CP properties of Higgs boson interactions with vector bosons has been performed by the ATLAS collaboration using the Run 2 dataset \cite{ATLAS_CP_combination}. CP-violating interactions are probed in the $H\to\gamma\gamma$, $H\to\tau\tau$, $H\to ZZ^*$, $H\to WW^*$, and $W\!H,H\to\bar{b}b$ channels. The analyses exploit observables which are asymmetric around zero in the presence of BSM CP-odd components. Only the shape of the distribution is used to constrain CP-violating effects.
The results are interpreted in the Standard Model Effective Field Theory (SMEFT) formalism, in the Warsaw basis.
Constraints on the CP-violating parameters \cHW, \cHB,  and \cHWB\ are derived considering either linear only or both linear and quadratic terms at dimension-six.

\begin{figure}[!htb]
    \centering
       \includegraphics[width=0.45\textwidth]{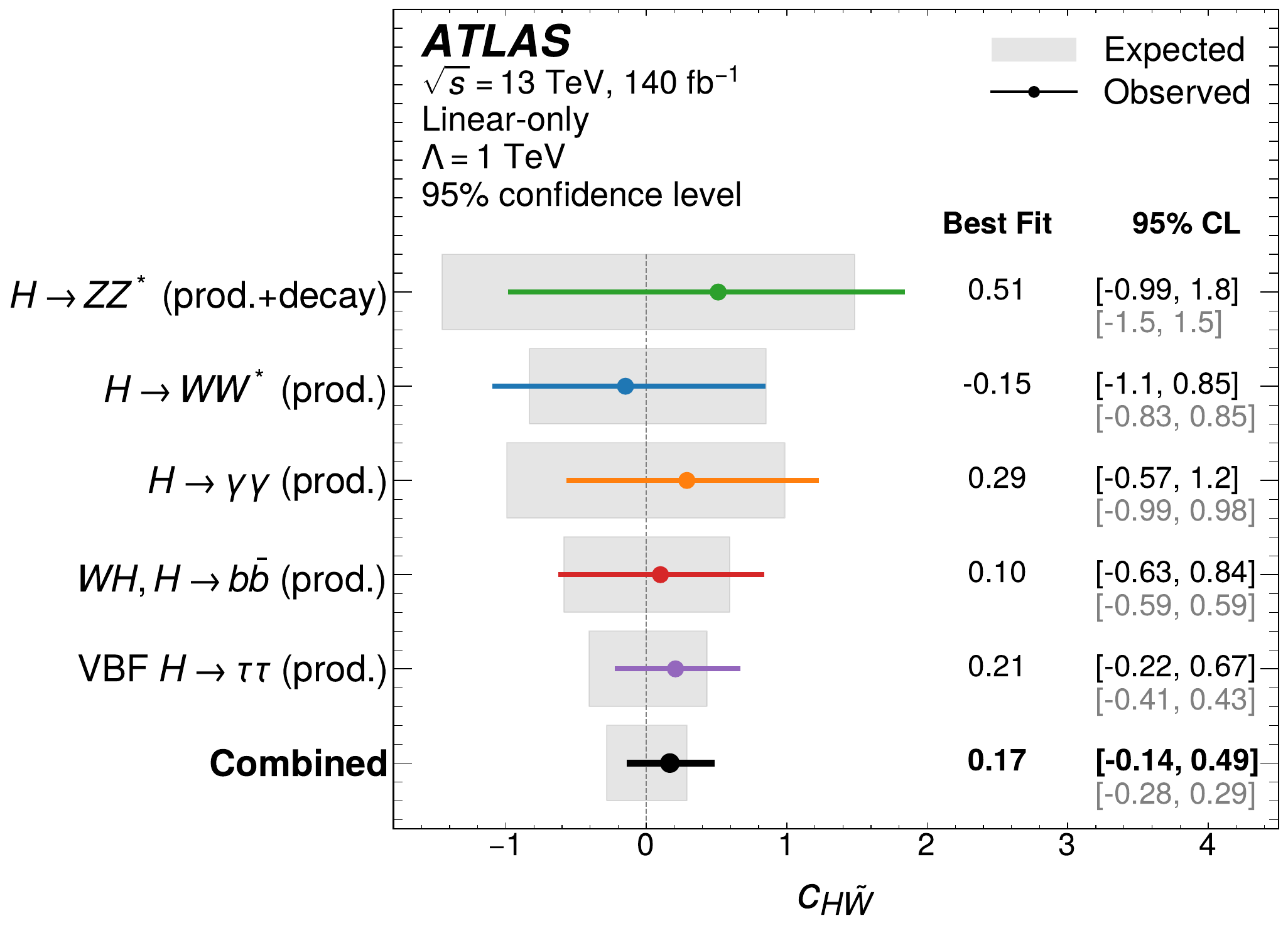}
       \includegraphics[width=0.44\textwidth]{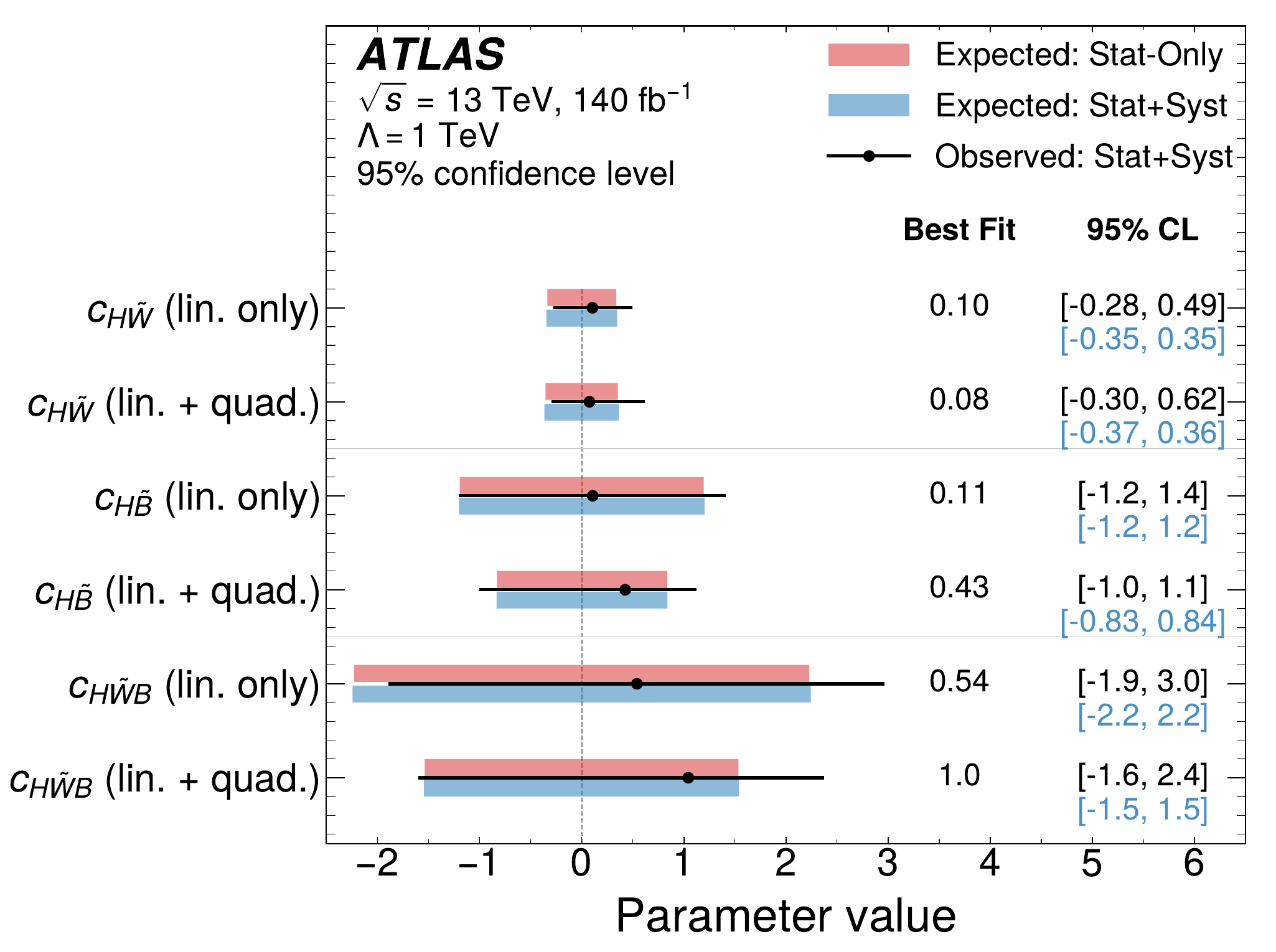}
    \caption{Left: Best-fit values and 95\% CL intervals on the \cHW\ operator for single-parameter fits ($\cHWB = \cHB = 0$) in each channel and the combination (considering linear only terms). Right: Best-fit values and 95\% CL intervals for the \cHW, \cHB, and \cHWB\ operators in a simultaneous fits to all three Wilson coefficients considering either linear only or both linear and quadratic terms. }
\label{fig:ATLAS_CP_cmb}
\end{figure}

Single-parameter fits for \cHW\ are performed, fixing $\cHWB = \cHB = 0$. The best-fit values and 95\% CL intervals considering linear terms only are shown in Fig.~\ref{fig:ATLAS_CP_cmb} (left) for individual channels and the combination. The combined observed 95\% CL interval for \cHW\ is $[-0.14,0.49]$. The limits are improved by more than 40\% compared to the most sensitive individual channel, $H\to\tau\tau$.
Results derived considering both linear and quadratic terms yield compatible values.
Simultaneous fits of all three CP-violating operators (\cHW, \cHB\  and \cHWB) are also performed, shown in Fig.~\ref{fig:ATLAS_CP_cmb} (right). 
The uncertainties on these results are strongly statistically limited. No evidence of CP violation is observed.

\subsection{Latest measurements of Higgs-vector boson couplings}

A study of the CP-nature of the Higgs-vector coupling was performed by the ATLAS collaboration considering Run 3 data \cite{ATLAS_CP_gg}. A matrix element-based optimal-observable technique is used to constrain \cHW. 
The observed value is $\cHW=0.24$ with a 95\% CL interval of $[-0.35,0.88]$, in good agreement with the SM. Compared with the previous Run 2 result, this represents a significant improvement in sensitivity of 38\%, due to improved machine learning techniques.
The combination of the two results leads to an observed value of $\cHW=0.25$ with a 95\% CL interval of $[-0.23,0.75]$.
An additional measurement probing Higgs couplings to longitudinally and transversely polarised vector bosons was performed, constraining polarisation dependent coupling strength scale factors ($a_L$ and $a_T$), expected to be unity in the SM. The observed values are $a_L=1.04$ $[0.93,1.17]$  (95\% CL) and $a_T=0.91$ $[0.70,1.14]$  (95\% CL) for fits considering both shape and rate effects.

A simultaneous measurement of eight Higgs-vector boson couplings in $H \to ZZ^*$ was performed by CMS considering Run 2 and Run 3 datasets \cite{CMS_Z_spin}.
Kinematic discriminants are used to derive constraints within the anomalous coupling framework. The best-fit value for the fractional contribution of CP-violating terms is $f_{a3}=0.16$ with a 95\% CL interval of $[0.01, 0.50]$. This deviation is linked to a data-asymmetry in the fitted kinematic observable, and remains statistically consistent with the SM due to the large parameter space explored.

\subsection{CP structure of the Yukawa coupling to tau leptons}

\begin{figure}[!htb]
    \centering
       \includegraphics[width=0.46\textwidth]{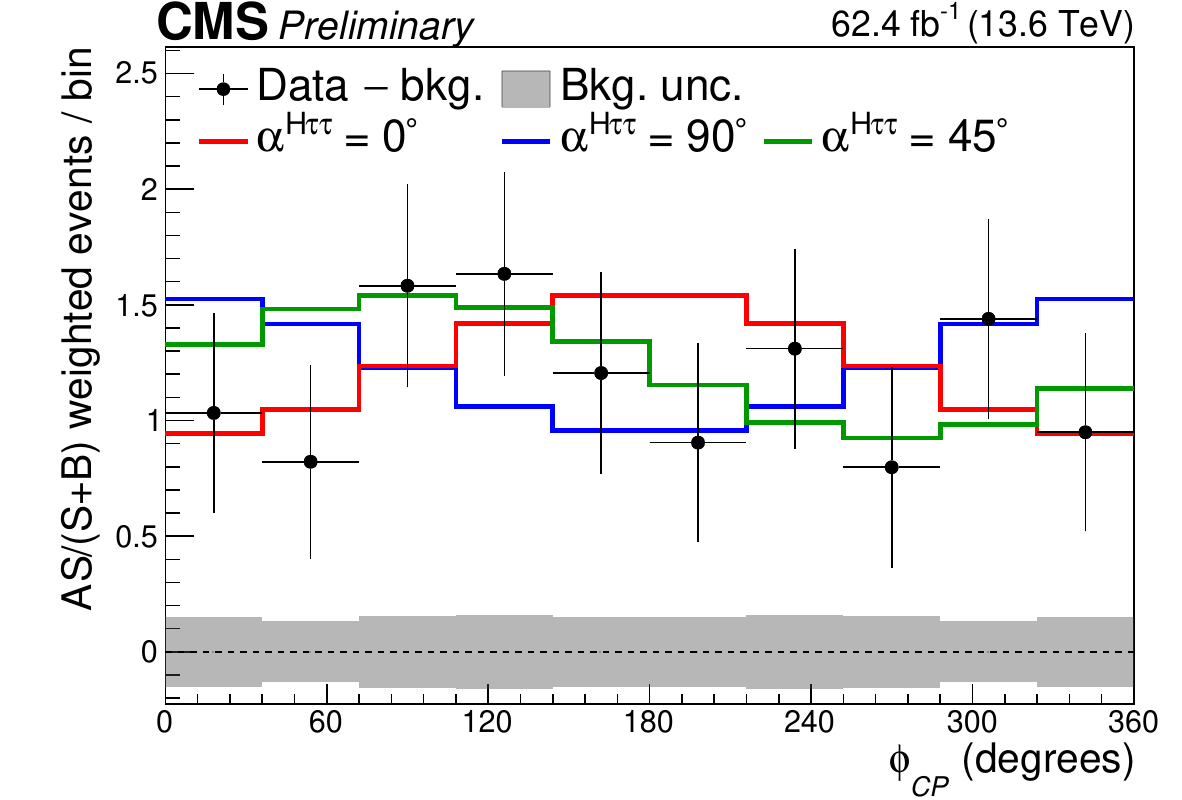}
       \includegraphics[width=0.46\textwidth]{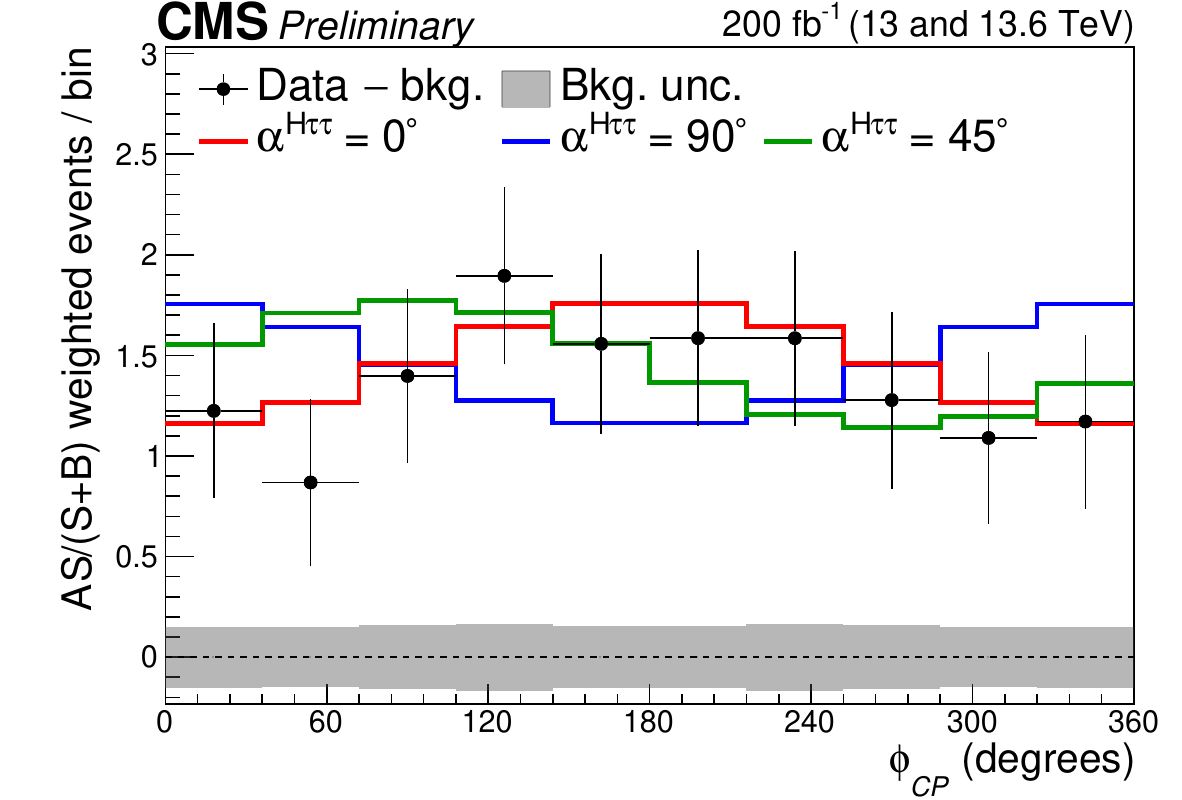}
    \caption{The \phicp\ distribution for early Run 3 (left) and the combination with the Run 2 result (right). 
    The background is subtracted from the data. 
    The CP-even distribution is depicted in red, the CP-odd displayed in blue, and a CP-mixed distribution shown in green.}
\label{fig:CMS_tau_CP}
\end{figure}

The CP nature of the Yukawa coupling between the Higgs boson and tau leptons has been measured by the CMS collaboration using Run 3 data \cite{CMS_tau_CP}.
Angular correlations between the decay products of tau leptons produced in $H\to\tau\tau$ decays are exploited to constrain the effective CP
mixing angle \alphacp, which parameterizes the admixture of scalar and pseudoscalar couplings. A non-zero value of \alphacp\ would indicate CP-violation. 
The mixing angle \alphacp\ is determined from the acoplanarity \phicp, defined as the angle between tau lepton decay planes in the Higgs boson rest frame.
The background-subtracted data and the expectation for different \alphacp\ scenarios is displayed in Fig.~\ref{fig:CMS_tau_CP} (left). 
The observed mixing angle is $\alphacp = 36^{+33}_{-30}{}^\circ$ compared with an expected value of $0\pm19^\circ$ under the SM hypothesis.
The Run 3 data show a slight preference for a CP-violating scenario.
The expected sensitivity is improved compared to the previous CMS result despite using less than half the amount of data.

This result is combined with the previous Run 2 CMS measurement.
The \phicp\ distribution for the combination is shown in Fig.~\ref{fig:CMS_tau_CP} (right). The data are more compatible with the CP-even distribution.
The observed value of \alphacp\ is $7\pm16^\circ$ ($0\pm14^\circ$ expected) with the uncertainty dominated by the statistical component. This represents the most precise measurement by CMS of \alphacp, with an expected precision that is the best achieved by any experiment to date.

\section{Summary}

An overview of recent measurements of Higgs boson properties using proton-proton collision data collected by the ATLAS and CMS experiments at the LHC during Run 2 and Run 3 has been presented. These include updated results on the Higgs boson mass and width, as well as constraints on the CP structure of Higgs boson couplings to vector bosons and to tau leptons.
No significant deviations from SM expectations are observed.
Many results remain limited by statistical uncertainties, and further improvements are expected with additional Run 3 data.

\end{document}